\documentclass[onecolumn,5p]{elsarticle}
\usepackage{graphicx}
\usepackage{amsmath,amssymb,bm}
\usepackage{color}

\begin{document}

\title{Warm baryogenesis}

\author[gra]{Mar Bastero-Gil}
\ead{mbg@ugr.es}

\author[edi]{Arjun Berera}
\ead{ab@ph.ed.ac.uk}

\author[br]{Rudnei O. Ramos}
\ead{rudnei@uerj.br}

\author[edi]{Jo\~ao G. Rosa}
\ead{joao.rosa@ed.ac.uk} 

\address[gra]{Departamento de F\'{\i}sica Te\'orica y del Cosmos, 
Universidad de Granada, Granada-18071, Spain}
\address[edi]{SUPA, School of Physics and Astronomy, University of Edinburgh, 
Edinburgh, EH9 3JZ, United Kingdom}
\address[br]{Departamento de F\'isica Te\'orica, Universidade do Estado do 
Rio de Janeiro, 
20550-013 Rio de Janeiro, RJ, Brazil}

\begin{abstract}

We show that a baryon asymmetry can be generated by dissipative effects during warm inflation via a supersymmetric two-stage
mechanism, where the inflaton is coupled to heavy mediator fields that then decay into light species through $B$- and $CP$-violating
interactions. In contrast with thermal GUT baryogenesis models, the temperature during inflation is always below the heavy mass threshold, simultaneously suppressing thermal and quantum corrections to the inflaton potential and the production of dangerous GUT relics. This naturally gives a small baryon asymmetry close to the observed value, although parametrically larger values may be diluted after inflation along with any gravitino overabundance. Furthermore, this process yields baryon isocurvature perturbations within the range of future experiments, making this an attractive and testable model of GUT baryogenesis. 
 
\end{abstract}




\maketitle


The inflationary paradigm \cite{inflation} has been extremely successful in explaining the flatness and homogeneity of the
observable universe, as well as providing an origin for Cosmic Microwave Background (CMB) anisotropies and the Large Scale
Structure (LSS). While models with a single slow-rolling scalar field have been extensively explored in the literature, these haven proven hard to embed within UV-completions of the Standard Model (SM) such as supergravity/string theory constructions, which typically suffer from the so-called `eta-problem' that precludes sufficiently long periods of inflation (see e.g.~\cite{Baumann:2009ni}).  

Warm inflation \cite{wi, BasteroGil:2009ec} (see also \cite{earlydissp}) offers an attractive solution to this problem by taking into account dissipative effects, which not only damp the inflaton's motion and allow for longer periods of slow-roll but also act as a source of light particles that may lead to a `graceful exit' into a radiation-dominated era. Furthermore, when the temperature of the  radiation exceeds the Hubble rate during inflation, $T>H$, thermal inflaton fluctuations become the main source of density perturbations, typically yielding lower inflationary scales than {\it cold} scenarios. 

{}From a supersymmetric (SUSY) two-stage mechanism \cite{BasteroGil:2009ec,Berera:2002sp}, in which interactions between the inflaton and the light particles are mediated by heavy `catalyzer' fields $X$, whose masses are {\it above} the temperature of the
radiation bath, $T\lesssim m_X/100$, the flatness of the inflaton potential is safe from both quantum and thermal corrections. As
supersymmetry is, however, inefficient in cancelling time non-local processes, this allows for strong dissipative effects with moderately large field multiplicities \cite{BasteroGil:2009ec}.  

Dissipation is naturally an out-of-equilibrium process, as annihilation of the resulting particles cannot efficiently `re-populate' the classical background condensate. It is then natural to ask whether a baryon asymmetry may be produced during inflation through dissipative effects, by incorporating $B$- and $CP$-violating interactions in the two-stage mechanism described above, thus 
satisfying the Sakharov conditions \cite{Sakharov:1967dj}. The structure of the two-stage interactions then suggests a parallel with 
thermal baryogenesis models in GUT scenarios, in which the $B$- and $CP$-violating decays of heavy GUT bosons may occur in an 
out-of-equilibrium fashion once the temperature of the universe drops below their mass threshold~\cite{Nanopoulos:1979gx}.  

In this Letter, we show that heavy GUT states can {\it mediate} dissipative processes during warm inflation leading to a baryon
asymmetry (see e.g. \cite{others} for related studies). An attractive feature of this mechanism is the generation of an asymmetry at $T\ll M_{GUT}$, avoiding the production of undesirable relics, like magnetic monopoles, that are generic in thermal scenarios. As we show below, the resulting asymmetry is naturally small in the low-temperature regime, despite the $\mathcal{O}(1)$ couplings typically required for successful warm inflation, as opposed to both thermal and non-thermal \cite{Kolb:1996jt} GUT baryogenesis scenarios where these are necessarily suppressed. 

We also show that the resulting  baryon-to-entropy ratio exhibits thermal fluctuations that lead to baryon isocurvature perturbations in 
the CMB spectrum and that may be accessible with upcoming experiments such as Planck \cite{planck}, a feature which is lacking in many of the proposed models of baryogenesis. {\it Warm baryogenesis} thus constitutes the first example of a consistent warm inflation scenario where a baryon asymmetry is produced and a viable model of GUT baryogenesis. 


We start by considering a GUT-like extension of the SUSY two-stage mechanism of \cite{Berera:2002sp} that includes interactions between
the inflaton, described by the scalar component $\phi$ of a superfield $\Phi$, and a set of mediator superfields $X_a$, $a=1,\cdots, N_X$. These are coupled to light degrees of freedom (dof) described by a set of $N_f$ superfields $Q_i$ and $L_i$, where the former carry a non-zero baryon number (or $B-L$ charge) $b_i$ and the latter correspond to non-baryonic species. {}For simplicity, we will
loosely refer to these as `quark' and `lepton' superfields, although our discussion applies to more generic scenarios. The relevant
superpotential can be written as: 
%
\begin{equation} \label{superpotential}
W=\bigg[g_a\Phi X_a^2+
  h_{a}^{ij}X_aQ_iQ_j+\lambda_a^{ij}X_aQ_i^cL_j\bigg]~,
\end{equation}
where a sum over the heavy and light field indices is implicit. As generic in SUSY GUT models, the form of the Yukawa terms in
Eq.~(\ref{superpotential}) is such that there is no consistent assignment of baryon number to any of the $X_a$ fields, which may then
decay in a $B$-violating fashion. The couplings $h_a^{ij}$ and $\lambda_a^{ij}$ are complex and $N_f\geq3$ in order to ensure
violation of $C$ and $CP$, whereas the $g_a$ couplings are real. The mass of the mediator supermultiplet components during inflation is
given by $m_a^2=2g_a^2\varphi^2$, where $\langle\phi\rangle=\varphi/\sqrt2$, with a negligible splitting between the fermionic and bosonic components due to soft SUSY breaking during inflation. We will also assume the masses of the $Q_i$ and $L_i$ components are well below the temperature during warm inflation and may thus be discarded.

As shown in the literature (see e.g. \cite{BasteroGil:2009ec}), the main processes contributing to dissipative effects in warm inflation
correspond to interactions between the scalar components in a superpotential of the form Eq.~(\ref{superpotential}), so that the
dominant source of a baryon asymmetry will be the production of squark fields. The relevant terms are included in the scalar
potential 
$V_{s}=4g_a^2|\phi|^2|\chi_a|^2+2g_a\phi^\dagger\chi_a^\dagger(h_a^{ij}\tilde{q}_i\tilde{q}_j+\lambda_a^{ij}\tilde{q}_i^\dagger
\tilde{l}_j)+\mathrm{h.c.}$,
where $\chi_a$, $\tilde{q}_i$ and $\tilde{l}_j$ denote the scalar components of the $X_a$, $Q_i$ and $L_j$ superfields, respectively. A baryon asymmetry will then arise from the difference between the particle production rates for scalar quarks and antiquarks. In \cite{Graham:2008vu}, these were computed in the adiabatic regime from the thermal Wightman self-energy for each light particle species $i$, yielding for the time derivative of the energy density,
%
\begin{equation} \label{ndot}
\dot{\rho}_i^{(d)}\!\!=\!\!\int \frac{d^3 p}{(2 \pi)^3}\, \omega_p 
\mathrm{Im}\bigg[2\int_{-\infty}^t
\!\! dt'  {e^{-i\omega_p(t-t')}\over2\omega_p}\Sigma_{21}(p,t,t')\bigg],
\end{equation}
where $p$ and $\omega_p$ are the momentum and energy of the light fields. The difference between this quantity for baryonic and anti-baryonic species then sets the rate at which a net baryon asymmetry is produced, and can be compard to the total particle production rate in order to determine the baryon-to-entropy ratio produced by warm inflation. Here we will outline the main steps of this computation, describing the more technical details in the appendix for the interested reader.

Firstly, one should note that in the low-temperature regime, where thermal corrections to the inflaton potential are suppressed, the main contribution to dissipative particle production comes from one-loop effects involving {\it virtual} heavy $\chi_a$ bosons, since production of on-shell states is Boltzmann suppressed for $m_a\gg T$. This is inherently different from most baryogenesis models in the literature involving the decay of on-shell heavy states, where the out-of-equilibrium condition results from the inbalance between direct decay and inverse decay processes \cite{Buchmuller:2000nd}. 

The leading particle production process is in this case \cite{Graham:2008vu} $\phi\rightarrow \sigma_i^2\sigma_j^2$, where $\sigma_{i,j}=q_{i,j}, l_{i,j}$, via virtual $\chi$ field pairs. Since the inflaton field is varying (slowly), the resulting particles cannot efficiently annihilate and fully give away their energy back into the background condensate. This leads to a net particle production and makes dissipative effects inherently out-of-equilibrium, as signaled by the non-zero value of the right-hand side of Eq.~(\ref{ndot}) for $\dot\varphi\neq 0$ (see appendix).

Although the leading contributions to the light particle self-energy arise at one-loop order, the associated diagrams correspond to `squared' tree-level diagrams, and hence cannot contribute to the baryon asymmetry according to the theorem by Nanopoulos and Weinberg
\cite{Nanopoulos:1979gx}. The diagrams contributing to the squark self-energy up to 2-loop order are then illustrated in Fig.~1, with
analogous diagrams for the anti-squark self-energy including the complex conjugate couplings. 
\begin{figure}[htbp]
\begin{center}
\includegraphics[scale=0.53]{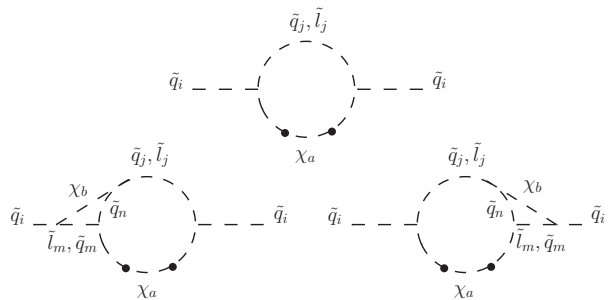}
\caption{Feynman diagrams contributing to the $\tilde{q}_i$
  self-energy up to 2-loop order. The filled circles correspond to
  couplings $2g_a^2\varphi^2$ with the inflaton background field.}
\end{center}
\end{figure}
We have computed these contributions in the low-temperature regime, $T\ll m_a$, justifying a low-momentum approximation for the heavy
boson propagators~\cite{Graham:2008vu}. Note that the $\chi_a$ propagators are `dressed', i.e.~involve a resummation over light field loop corrections. These correspond to the non-vanishing decay width of the heavy bosons, $\Gamma_{\chi_a}$, which induces particle production despite their virtual nature in the low-temperature regime (see appendix). We then obtain for the difference between the squark and anti-squark self-energies:
%
\begin{equation} \label{self_energy_diff}
\Delta\Sigma_{21}^{B}=8\sum_{a,b=1}^{N_X} \int
d^3k~f_am_b^2I_{ab}\mathrm{Im}\big(\mathrm{Tr}[\bm{\lambda}_b^*
\bm{\lambda}_a^T\mathbf{h}_b\mathbf{h}_a^\dagger]\big)~,
\end{equation}
where $k$ denotes the momentum of the $\chi_a$ propagators in the main loop, $f_a(p,k,t,t')$ includes the main loop factors and is explicitly given in the appendix, while $I_{ab}(p,k)$ is the imaginary part of the `triangle' loop integrals. In the low-temperature regime for the heavy particles, a low-momentum approximation applies and we obtain %
\begin{equation} \label{triangle_loop}
m_b^2I_{ab}(p,k)\simeq-{1\over16\pi}\bigg(1-{1\over2}{p^2-(p-k)^2\over m_b^2}\bigg)~.
\end{equation}
{}From Eqs. (\ref{self_energy_diff}) and (\ref{triangle_loop}), it is easy to conclude that, as in thermal GUT baryogenesis models
\cite{Nanopoulos:1979gx}, one needs at least two distinct heavy fields, with either different masses or different Yukawa couplings, in
order to produce a non-vanishing baryon asymmetry. In particular, note that the trace $\mathrm{Tr}[\bm{\lambda}_b^*\bm{\lambda}_a^T\mathbf{h}_b\mathbf{h}_a^\dagger]$ is real for $a=b$ and that the $I_{ab}$ and $I_{ba}$ contributions cancel each other unless $m_a\neq m_b$.

The total particle production rate can be computed by adding the self-energy of all the light baryonic and non-baryonic fields, for
which it is sufficient to consider the lowest-order diagrams in Fig.~1. This gives:
%
\begin{equation} \label{self_energy_sum}
\Sigma_{21}^{R}=2\int
d^3k~f_a\bigg[\mathrm{Tr}[\mathbf{h}_a\mathbf{h}_a^\dagger]+
\mathrm{Tr}[\bm{\lambda}_a^*\bm{\lambda}_a^T]+
\mathrm{Tr}[\bm{\lambda}_a\bm{\lambda}_a^\dagger]\bigg]~,
\end{equation}
where the first two terms correspond to the self-energy of the light baryonic species and the last term to the non-baryonic fields. The
internal and external momentum integrations, as well as the time integration in Eq.~(\ref{ndot}), can then be performed following the
procedure in \cite{Graham:2008vu}. The momentum dependence of the triangle loop integrals $I_{ab}$ yields an additional factor $T^2/m_b^2$ with respect to the leading order result, while changing the overall value by a factor $\sim3.5$, which we obtained numerically (see e.g. \cite{dissippaper}). This corresponds to an inherent momentum cut-off in Eq.~(\ref{triangle_loop}) due to Boltzmann suppression of the heavy mediators in the low-$T$ regime, such that $p,k\lesssim T$. 

{}For concreteness, let us consider a case with $N_X-1$ degenerate fields of mass $m_X$ and an additional multiplet of mass $m_X+\Delta m_X$, with $\Delta m_X/m_X\ll1$. We also assume without loss of generality that all $Q_i$ superfields carry the same baryonic charge, $b$, and that the Yukawa couplings only differ by a phase, i.e.~$\bm{h}_a=|h|\bm{V}e^{i\theta_a}$ and
$\bm{\lambda_a}=|\lambda|\bm{U}e^{i\alpha_a}$, with $\bm{U}$ and $\bm{V}$ denoting unitary matrices. We find the asymmetry ratio to be 
%
\begin{equation} \label{ratio}
r\equiv{\dot{\rho}_B^{(d)}\over
  \dot{\rho}_R^{(d)}}\approx 3.5 {b\, \sin\delta\over4\pi}{|h|^2|\lambda|^2\over
  |h|^2+2|\lambda|^2}{N_X-1\over N_X}{T^2\over m_X^2}{\Delta m_X\over
  m_X},\!\!\!
\end{equation}
where $\delta$ denotes an effective CP-violating phase. Note that this is independent of the number of light species $N_f$ and only mildly dependent on $N_X$.

During inflation, the radiation density is determined by both the dissipative effects and the Hubble friction, with evolution equation 
$\dot\rho_R+4H\rho_R=\Upsilon\dot\varphi^2$. The dissipative coefficient, computed in the low-$T$ regime~\cite{Graham:2008vu,dissippaper,BasteroGil:2009ec}, is of the  form $\Upsilon=C_\phi T^3/\varphi^2$, where $C_\phi\simeq
0.16|h|^4N_XN_{decay}^2$ for $|h|\simeq|\lambda|$, with $N_{decay}$ denoting the number of available decay channels for the 
heavy mediators. Similarly, a fraction $r\Upsilon$ sources a net baryon number density $n_B=n_b-n_{\bar{b}}$. Since 
$\rho_R=(g_* \pi^2/30)T^4$, with $g_*$ denoting the number of relativistic dof, the entropy and baryon number densities are given by: 
%
\begin{eqnarray} \label{entropy}
 \dot{s}+3Hs&=&{\Upsilon\dot\varphi^2\over
   T}~,\nonumber\\
\dot{n_B}+3Hn_B&=&{{45 \zeta(3) }\over{2 \pi^4}}{g_B \over g_*}
r{\Upsilon\dot\varphi^2\over T},
\end{eqnarray}
with $g_B$ giving the number of baryonic dof. Note that this includes fermionic fields, which are thermally produced despite the dominant scalar nature of the dissipative processes~\cite{dissippaper}. In the slow-roll regime, these quantities quickly approach a steady state solution, yielding for the baryon-to-entropy ratio $\eta_s= n_B/s$, 
%
\begin{equation} \label{baryon_asymmetry}
\eta_s\approx \! 3.5 {45\zeta(3)\over 8\pi^5} {g_B b\, \sin \delta \over g_*}
{|h|^2|\lambda|^2\over
  |h|^2+2|\lambda|^2}{N_X-1\over N_X}{T^2\over m_X^2}{\Delta m_X\over
  m_X}.
\end{equation}
In warm inflation models, radiation typically comes to dominate the energy density \cite{wi}, ending inflation during the
slow-roll regime, such that Eq.~(\ref{baryon_asymmetry}) gives the observed value if no significant entropy
production or washout processes occur in the post-inflationary evolution. In particular, one can ensure that electroweak sphaleron
processes \cite{Klinkhamer:1984di} conserve the produced asymmetry by generating a net $B-L$ number density during warm inflation. Also, if the `catalyzer' fields are heavy (Majorana) right-handed neutrinos with $L$-violating interactions
\cite{Fukugita:1986hr}, an inflationary lepton asymmetry may later be converted into baryon number by sphaleron processes.

Apart from numerical factors, the main difference between the baryon asymmetry produced by dissipation and the corresponding result for out-of-equilibrium decay of heavy bosons resides in the $T^2/m_X^2$ suppression in the low-$T$ regime, so that the heavy fields are not
excited by background dissipation. This is an appealing feature, since it not only prevents the generation of GUT relics that could overclose the universe but also leads to a small baryon-to-entropy ratio, at the same time ensuring that thermal corrections are under control. Taking for instance $b=1/3$, $|h|\simeq |\lambda|$, $N_X\gg1$ and the MSSM values for $g_B$ and $g_*$, we then obtain
%
\begin{equation} \label{baryon_asymmetry_estimate}
\eta_s\approx 8.9\times10^{-11}|h|^2\bigg({T/m_X\over 0.01}\bigg)^2\bigg({\frac{\Delta
    m_X}{m_X}\over 0.015}\bigg)\bigg({\sin\delta\over0.025}\bigg).
\end{equation}
Note that the moderately large field multiplicity typically required in warm inflation models will not affect the ratio $g_B/g_*$ significantly. The best present estimate for the baryonic asymmetry comes from Big-Bang nucleosynthesis~\cite{etaBBN}, 
$7.2  \times 10^{-11} \leq \eta_s \leq 9.2 \times 10^{-11}$, at 95$\%$ C.L. Our result (\ref{baryon_asymmetry_estimate}) may thus be naturally within the observed window despite the $\mathcal{O}(1)$ couplings typically required in warm inflation scenarios, with mass degeneracy and a CP-violating phase at only the few percent level. On the other hand, thermal GUT baryogenesis scenarios require strongly suppressed couplings in order to reproduce the observed asymmetry. Similarly, in non-thermal scenarios \cite{Kolb:1996jt}, the parametric resonance that excites the heavy bosons is efficient only for sufficiently long lifetimes.

In some warm inflationary scenarios, radiation is always subdominant even though dissipation sustains the required number of e-folds \cite{BasteroGil:2009ec}. One then expects additional entropy production through conventional reheating processes, as well as possibly from the late decay of any light moduli present during inflation. Notice, however, that moduli dynamics may be modified due to the thermal radiation bath with $T>H$. Although this may dilute any undesired gravitino abundance \cite{Sanchez:2010vj}, it will also dilute the asymmetry produced by dissipation, but this may be compensated with larger $CP$ phases, couplings and non-degeneracies in the heavy mediator spectrum.

 
Given that $m_X\propto\phi$, Eq.~(\ref{baryon_asymmetry}) also implies $\eta_s\propto (T/\phi)^2$, so that thermal fluctuations of the inflaton field will be imprinted on the baryon-to-entropy ratio. Although baryons are subdominant during inflation, they become a significant component of the energy density at late times and will contribute to CMB anisotropies and LSS. Although such baryon isocurvature perturbations (BIP) also arise in other scenarios \cite{isocurvature}, they are in this case fully (anti-)correlated with adiabatic perturbations, since both originate from inflaton fluctuations. These can be obtained by perturbing the inflaton equation, $\ddot\phi+(3H+\Upsilon)\dot\phi+V_\phi=0$, and are coupled to the temperature fluctuations via the evolution equation for $\rho_{R}$. BIP are conventionally measured by the ratio $B_B=S_B/\zeta$, where $\zeta=-H\delta\rho/\dot\rho$ is the gauge-invariant curvature perturbation and $S_B=\delta\rho_B/\rho_B-(3/4)\delta\rho_R/\rho_R=\delta\eta_s/\eta_s$ \cite{Lyth:2002my}. In the slow-roll regime, on superhorizon scales, we obtain after some algebra: 

\begin{eqnarray} \label{isocurvature_pert_ratio}
B_B={2\big[2\eta_\phi(1+Q)-\sigma_\phi(3+5Q)-\epsilon_\phi(3+Q)\big]\over(1+Q)^2(1+7Q)},
\end{eqnarray}
where we define $Q=\Upsilon/3H$, $\epsilon_\phi=(m_P^2/2)(V_\phi/V)^2$, $\eta_\phi=m_P^2V_{\phi\phi}/V$ and $\sigma_\phi=m_P^2V_\phi/(V\phi)$. This thus yields an additional observable that may be used to probe the consistency of warm inflation models, along with the spectral index, the tensor-to-scalar ratio and non-gaussianity parameters. In particular, we obtain \cite{BasteroGil:2009ec}
\begin{equation} \label{scalar_index}
{B_B\over n_S-1}\simeq
\begin{cases}
{3\epsilon_\phi-2\eta_\phi+3\sigma_\phi\over\epsilon_\phi-\sigma_\phi}~, & Q\ll1\\
{2\over3Q}{\epsilon_\phi-2\eta_\phi+5\sigma_\phi\over3\epsilon_\phi+\eta_\phi-6\sigma_\phi}~, & Q\gg1
\end{cases}
\end{equation}
so that $B_B$ is generically at most of the same order of magnitude as the deviations from scale invariance, with $n_S=0.968\pm0.012$ (68\% C.L.) \cite{Komatsu:2010fb}, being further suppressed for strong dissipation. For example, for a quartic potential one finds $B_B\simeq-0.096,-0.007/Q$ in the weak and strong dissipation regimes, respectively. The most recent WMAP analysis of cold dark matter anti-correlated isocurvature perturbations \cite{Komatsu:2010fb}, taking into account that $\Omega_c/\Omega_b\simeq 5$, yields $|B_B|<0.34$ (95\% C.L.), for $B_B<0$, while according to an earlier analysis $-0.53<B_B<0.43$ (95\% C.L.) \cite{Gordon:2002gv}. This is thus generically consistent with the expected amount of BIP, even for $Q\ll1$, while Planck should improve these bounds by an order of magnitude~\cite{planck}.


Generating a baryon asymmetry through dissipative effects during warm inflation thus exhibits several attractive features. By keeping the temperature during inflation below the GUT scale, it provides an appealing model of GUT baryogenesis, with no unwanted relics and a small baryon-to-entropy ratio with no unnaturally suppressed couplings and CP-violating angles, at the same time keeping thermal corrections to the inflaton potential under control. Furthermore, a superpotential of the form in Eq.~(\ref{superpotential}) is natural in SUSY GUT constructions where the inflaton lies within a diagonal subgroup of an adjoint Higgs representation and the heavy mediators correspond to its off-diagonal components, as in the D-brane model considered in \cite{BasteroGil:2011mr}. These constructions are also natural arenas for warm inflation due to the large field multiplicity available, ensuring sufficiently strong friction effects to sustain the required number of e-folds. 

Although in this work we have focused on the baryon asymmetry generated by the varying inflaton field, it is possible in some models of warm inflation for excitations above the background condensate, $\phi=\langle\phi\rangle+\phi_1$, to also play a significant role. The slow roll conditions require $\eta_\phi<1+Q$, where $Q=\Upsilon/3H$, which imply $m_\phi/H\lesssim \sqrt{1+Q}$. Since $T>H$ during warm inflation, one could have $m_\phi\lesssim T$ if $Q\ll1$ in the early stages of inflation, such that the process $\chi\rightarrow \sigma_i^2 \phi_1$ contributes to $\Gamma_{\chi}$ and hence the $\phi_1$ particle states become a non-negligible component of the radiation bath. They are, however, unstable and decay via $\phi_1\rightarrow \sigma_i^2\sigma_j^2$ through virtual $\chi$ fields. If dissipation becomes strong towards the end of inflation, $Q\gg1$, as is the case for typical inflationary potentials \cite{BasteroGil:2009ec}, it is possible to have $m_\phi>T$, depending on the evolution of $\varphi$ and $T$, so that the $\phi_1$ states decay out-of-equilibrium and produce a net baryon number. Given the common origin and structure of the relevant interactions, the contributions from background dissipation and $\phi_1$ decay to the baryon asymmetry will necessarily have the same sign and similar magnitude, so that our earlier results are robust despite the dependency of the latter contribution on the details of the inflationary dynamics.

One of the crucial aspects of the model presented in this Letter is the fact that dissipation provides a novel mechanism for satisfying the Sakharov out-of-equilibrium condition \cite{Sakharov:1967dj}. Although we have examined a nearly thermal equilibrium regime, warm inflation is general to any non-equilibrium inflationary scenario governed by fluctuation-dissipation dynamics \cite{wi}. Furthermore, dissipative processes may also play an important role in the post- inflationary universe, e.g.~during cosmological phase transitions, so that we may extrapolate from our analysis a more general mechanism of {\it dissipative baryogenesis}.

Warm baryogenesis thus provides an unprecedented link between inflationary physics and the observed baryon asymmetry, a connection that may be tested in the near future and possibly shed a new light on two of the most important problems in modern cosmology.


\section*{Acknowledgements}

M.~B.~G. is partially supported by MICINN (FIS2010-17395) and ``Junta de Andaluc\'ia'' (FQM101). A.~B.~and J.~G.~R.~are supported by STFC.  R.~O.~R.~is partially supported by CNPq (Brazil).


\appendix
\section{Particle production rates}

The scalar interactions involving the background inflaton field, $\varphi$, the mediators $\chi_a$ and the light fields $\sigma_i=q_i,l_i$ relevant for the dissipative dynamics are of the form (omitting the indices for simplicity):
\begin{equation} \label{Lphichisigma}
{\cal L}_I = - 2g^2\varphi^2 \chi^\dagger\chi - h M [\chi^\dagger\sigma^2 + \chi (\sigma^\dagger)^2 ] ~,
\end{equation}
where $M=\sqrt{2}g\varphi$. Following \cite{Graham:2008vu}, the particle production rate of $\sigma$-particles in the radiation bath due to interactions with the $\chi$-fields is given by:
\begin{equation} \label{dotn}
\dot{n}_\sigma(p) = {\rm Im} \left[ 2 \int_{-\infty}^t dt' \frac{e^{-i \omega({\bf p})(t-t')}}{2 \omega({\bf p})} \Sigma_{21} ({\bf p},t,t')\right]~,
\end{equation}
with the Wightman self-energy given, to leading order, by:
\begin{eqnarray} \label{Sigma12}
\!\!\!\!\!\!\!\!\!\!\!\!\!\!\!\Sigma_{\sigma,21} ({\bf p}, t, t') &=& 16g^4 h^2 M^2 \int \frac{d^3 k}{(2 \pi)^3}
\prod_{i=1}^3\frac{d \omega_i}{2 \pi}
e^{-i \omega_2(t-t')} \nonumber\\
& & \!\!\!\!\!\!\!\!\!\!\!\!\!\!\!\!  \times G_{\sigma,21}({\bf p} - {\bf k}, t-t') G_{\chi,2}^a({\bf k}, \omega_1)  \varphi^2 (\omega_1-\omega_3)\nonumber\\
& & \!\!\!\!\!\!\!\!\!\!\!\!\!\!\!\! \times  G_{\chi,a}^b({\bf k}, \omega_3)  \varphi^2 (\omega_3-\omega_2)G_{\chi,b\,1}({\bf k}, \omega_1) ~,
\end{eqnarray}
where $ G_{\chi,a\,b}({\bf k}, \omega)$, $a,b=1,2$, are the fullly dressed finite temperature Schwinger-Keldysh $\chi$-propagators. In the low-temperature regime, $T\ll m_\chi$, the main contributions to this integral arise from virtual low-momentum modes, so that we can use the approximate forms:
\begin{eqnarray}
G_{\chi,a}^a({\bf k},\omega)&\approx& - \frac{i}{m_\chi^2}~,\nonumber\\
\nonumber \\
G_{\chi,2}^1({\bf k}, \omega) &\approx& \frac{2 \Gamma_\chi({\bf k},\omega)}
{m_\chi^3} \left[ 1 + n(\omega)\right]~,
\label{Gab}
\end{eqnarray} 
where $n(\omega)$ is the Bose-Einstein distribution function and $\Gamma_\chi$ is the finite temperature width of the $\chi$-field, which has been computed in \cite{dissippaper} at one-loop order. In the slow-roll regime one can approximate the Fourier transform of the background inflaton field $\varphi^2(\omega)\simeq 2i \varphi \dot{\varphi}2\pi\delta'(\omega)$. Also, the light $\sigma$-propagators can be Fourier transformed to yield:
\begin{eqnarray} \label{Gsigma12}
\!\!\!\!\!\!\!\!\!\!\!\!\!\!\!G_{\sigma,21}({\bf p}-{\bf k},\omega) &=& -i \left[1 + n(\omega)\right]\frac{2 \pi}{2\omega^\sigma_{{\bf p}-{\bf k}}}
\big[\delta(\omega-\omega^\sigma_{{\bf p}-{\bf k}})  + \nonumber\\
&+&\delta(\omega+\omega^\sigma_{{\bf p}-{\bf k}})\big]{\rm sgn}(\omega)~,
\end{eqnarray}
where ${\rm sgn}(\omega)$ is the sign function. Substituting these results into Eq.~(\ref{Sigma12}), we get:
\begin{eqnarray} \label{Sigma12-3}
\Sigma_{\sigma,21} ({\bf p}, t, t') &=& h^2\int d^3k\ f(p,k,t,t')~,
\end{eqnarray}
where
\begin{eqnarray} \label{function_f}
f(p,k,t,t') &=& -{64i g^4\over (2 \pi)^3}\frac{M^2 \varphi^2 \dot{\varphi}^2}
{m_\chi^7} \int\frac{d \omega_2}{2 \pi} \frac{1}{\omega^\sigma_{{\bf p} - {\bf k}}}\nonumber\\
& \times& \left\{ \Gamma_{\chi}({\bf k},\omega_2) \left[ 1 + n(\omega_2)\right]\right\}''\nonumber\\
& \times&\left\{    e^{-i (\omega_2+\omega^\sigma_{{\bf p} - {\bf k}})(t-t')}
\left[1+n(\omega^\sigma_{{\bf p} - {\bf k}}) \right] \right. \nonumber\\
& & + \left. e^{-i (\omega_2-\omega^\sigma_{{\bf p} - {\bf k}})(t-t')}
n(\omega^\sigma_{{\bf p} - {\bf k}}) \right\} 
\end{eqnarray}
corresponds to the function appearing in Eqs.~(\ref{self_energy_diff}) and (\ref{self_energy_sum}). To obtain the net particle production rate it remains to perform the time integration in Eq.~(\ref{dotn}), for which we may use the following identity:
\begin{eqnarray} \label{identity}
\int_{-\infty}^t dt'& \!\!\!\!\!\!\!\!\!\!\!\!\!\!\!\!\!\!\!\!\!\!\!\!\!\!\!\!\!\!\!\!\!\!\!\!\!\!\!\!\!\!\! e^{-i (\omega_2+\omega^\sigma_{{\bf p}} \pm \omega^\sigma_{{\bf p} - {\bf k}})(t-t')} =\nonumber\\
&\pi \delta( \omega_2+\omega^\sigma_{{\bf p}} \pm \omega^\sigma_{{\bf p} - {\bf k}}) +
\frac{i}{ \omega_2+\omega^\sigma_{{\bf p}} \pm 
\omega^\sigma_{{\bf p} - {\bf k}}} ~.
\end{eqnarray}
This yields:
\begin{eqnarray} \label{dotn-2}
\!\!\!\!\!\!\!\!\!\!\!\!\!\dot{n}_\sigma(p) &=& -32 g^4 h^2 M^2 \frac{\varphi^2 \dot{\varphi}^2 }{m_\chi^7}
\int \frac{d^3 k}{(2 \pi)^3} \frac{1}{\omega^\sigma_{{\bf p}}
\omega^\sigma_{{\bf p} - {\bf k}} } \times\nonumber\\
& &
\!\!\!\!\!\!\!\!\!\!\!\!\bigg[\left\{ \Gamma_{\chi}({\bf k}, -\omega^\sigma_{{\bf p}} - 
\omega^\sigma_{{\bf p} - {\bf k}} ) \left[ 1 + n( -\omega^\sigma_{{\bf p}} - 
\omega^\sigma_{{\bf p} - {\bf k}}  )\right]\right\}''\nonumber\\ 
& &\times\left[1 + n( \omega^\sigma_{{\bf p} - {\bf k}} ) \right] +. 
\nonumber \\
&+& 
\!\!\!\!\left.\left\{ \Gamma_{\chi}({\bf k}, -\omega^\sigma_{{\bf p}} + 
\omega^\sigma_{{\bf p} - {\bf k}} ) \left[ 1 + n( -\omega^\sigma_{{\bf p}} + 
\omega^\sigma_{{\bf p} - {\bf k}}  )\right]\right\}''\right.\nonumber\\ 
& & \times n( \omega^\sigma_{{\bf p} - {\bf k}} ) \bigg]~.
\end{eqnarray}

The total particle production rate in an expanding universe $\dot\rho_\sigma+4H\rho_\sigma=\Upsilon\dot\varphi^2$ can then be obtained by integrating over the momenta of the produced $\sigma$-particles: 
\begin{equation} \label{Upsilon}
\Upsilon = \frac{1}{\dot{\varphi}^2} \int \frac{d^3 p}{(2 \pi)^3}
\omega^\sigma_{{\bf p}} \dot{n}_\sigma(p)~.
\end{equation}
These results can then be used to compute the production rate of each species. For the difference between squark and antisquark production rates one needs to include the additional momentum dependence and the associated coupling structure arising from the triangle loop diagrams in the above expressions, according to Eq.~(\ref{triangle_loop}), while for the total particle production rate it suffices to add the one-loop contributions of all species, with the relevant couplings.


\end{document}